\documentclass[conference]{IEEEtran}
\IEEEoverridecommandlockouts

\usepackage{cite}
\usepackage{amsmath,amssymb,amsfonts}
\usepackage{algorithmic}
\usepackage{graphicx}
\usepackage{textcomp}
\usepackage{xcolor}
\usepackage{url}
\usepackage[utf8]{inputenc}
\usepackage{listings}

\definecolor{codegray}{rgb}{0.5,0.5,0.5}
\definecolor{light-gray}{gray}{0.95} 

\lstdefinestyle{ieee_verilog}{
    commentstyle=\color{codegray}\itshape,
    keywordstyle=\ttfamily\footnotesize,
    basicstyle=\ttfamily\footnotesize,
    breakatwhitespace=false,         
    breaklines=true,
    captionpos=b,                    
    keepspaces=true,
    aboveskip=1pt,                  
    belowskip=1pt,                  
    showspaces=false,                
    showstringspaces=false,
    showtabs=false,                  
    tabsize=2,
    language=Verilog,
    frame=tb,
    rulecolor=\color{black}
}

\lstdefinestyle{terminal_output}{
    backgroundcolor=\color{light-gray}, 
    basicstyle=\ttfamily\scriptsize,  
    caption=false,
    aboveskip=1pt,                  
    belowskip=1pt,
    frame=single,                       
    rulecolor=\color{black},
    breaklines=false,
}

\def\BibTeX{{\rm B\kern-.05em{\sc i\kern-.025em b}\kern-.08em
    T\kern-.1667em\lower.7ex\hbox{E}\kern-.125emX}}
    
\begin{document}

\title{Physical Design of UET-RVMCU: A Streamlined Open-Source RISC-V Microcontroller
\thanks{This work was conducted as part of the undergraduate research at the University of Engineering and Technology, Lahore.}
}

\author{\IEEEauthorblockN{Abdullah Azhar, Uneeb Kamal, Wajid Ali, Saad Gillani, Dr. Suleman Sami Qazi}
\IEEEauthorblockA{\textit{Department of Electrical Engineering} \\
\textit{University of Engineering and Technology, Lahore, Pakistan} \\
\{2021ee61, 2021ee81, 2021ee79, 2021ee118\}@student.uet.edu.pk, suleman.qazi@uet.edu.pk}
}

\maketitle

\begin{abstract}
This paper presents the design and physical implementation of UET-RVMCU, a lightweight RISC-V microcontroller derived from the UETRV-PCore. Aimed at creating an accessible and flexible open-source RISC-V-based microcontroller, UET-RVMCU simplifies the application-class UETRV-PCore by reducing pipeline stages, removing MMU functionality, and integrating GPIO peripherals. The final GDSII layout was generated using an open-source RTL-to-GDS flow (OpenLane). This project demonstrates the feasibility of transforming an application-class SoC into a feature-rich microcontroller suitable for embedded systems, emphasizing low area, design simplicity, and open-source development.
\end{abstract}


\section{Introduction}
The evolution of computing demands increasingly efficient, customizable, and cost-effective hardware solutions. Traditional proprietary instruction set architectures (ISAs) often impose limitations due to licensing costs and restricted accessibility. In contrast, RISC-V emerges as a transformative open-source ISA, offering a modular and extensible framework that empowers designers to develop tailored processors across a spectrum of applications—from embedded systems to high-performance computing.

RISC-V’s open nature fosters innovation by eliminating licensing barriers, thereby accelerating research and development in both academia and industry. Its simplified design principles facilitate easier implementation and verification processes, making it an ideal choice for educational purposes and experimental architectures.

Leveraging the advantages of RISC-V, the UET-RVMCU project aims to design and implement a streamlined 32-bit microcontroller unit (MCU) tailored for educational and research applications. This MCU integrates essential peripherals and adheres to the RV32IMA instruction set, balancing performance with resource efficiency. The project’s objective is to provide a practical platform for exploring processor design, embedded system development, and hardware-software co-design methodologies.

To realize the physical design of the UET-RVMCU, we employ OpenLane, an open-source digital ASIC implementation flow built upon tools primarily from the OpenROAD, YosysHQ, and Open Circuit Design projects. OpenLane defines a complete RTL-to-GDSII flow by integrating these tools with a series of custom scripts and utilities that address compatibility and methodology gaps. Initially, OpenROAD tools operated independently, each with its own infrastructure and interfaces. While this modular design followed the UNIX philosophy of doing one thing well, the fragmentation increased I/O overhead during iterative passes, reducing overall efficiency.

To address this, a shared physical data model OpenDB was introduced, enabling seamless data exchange between tools and significantly reducing file I/O costs. This unification allowed the development of custom tools within OpenLane that facilitate end-to-end automation of digital ASIC design. OpenLane supports two major workflows: hardening RTL designs into soft macros, and integrating those macros into a complete system-on-chip (SoC) layout. The flow has been successfully used to tape out a family of RISC-V-based SoCs known as striVe, demonstrating its reliability and capability for real-world chip fabrication.

This paper presents a comprehensive experience report on utilizing OpenLane for the physical design of the UET-RVMCU. We detail the end-to-end design flow, highlight challenges encountered during implementation, and discuss the solutions adopted to overcome them. Our findings aim to contribute to the collective knowledge of open-source hardware design and serve as a reference for future endeavors in the field. 

Macro hardening refers to the process of transforming a synthesized RTL design into a physical layout block (also called a "hardened macro") that can be integrated into a larger chip. In OpenLane, this involves going through the full ASIC implementation flow—including synthesis, floorplanning, placement, clock tree synthesis, routing, and signoff steps such as design rule checking (DRC) and layout versus schematic (LVS) verification. The resulting hardened macro is a standalone, manufacturable layout block (typically in GDSII format) with well-defined power, ground, and I/O boundaries.

This hardened macro can then be reused or instantiated in more complex SoCs, enabling hierarchical chip design. Macro hardening is especially useful for modular design approaches, where validated building blocks are integrated into larger systems.

\section{Architecture Overview of UET-RVMCU}

\subsection{Pipeline Reduction}
The UET-RVMCU adopts a simplified 3-stage pipeline architecture, consisting of Fetch, Decode/Execute, and Writeback stages. This reduction from a more complex pipeline structure decreases control logic overhead and silicon area, making it ideal for low-power, resource-constrained embedded environments. The streamlined pipeline also facilitates easier debugging and analysis for educational purposes.

\subsection{Privilege Modes}
In alignment with microcontroller requirements, the UET-RVMCU supports only the Machine (M) mode. The exclusion of Supervisor and User privilege modes reduces hardware complexity and software overhead, while still maintaining sufficient control and configurability for embedded system applications.

\subsection{Memory Management}
To further reduce design complexity and enhance determinism, the UET-RVMCU omits the Memory Management Unit (MMU) and virtual memory support. The processor operates under a flat memory model, which simplifies address translation and access control, thereby improving real-time responsiveness.

\subsection{Memory Architecture}
The memory system incorporates four block memories, each supporting byte-addressable access. This structure enables efficient separation of code and data memory while offering flexibility in memory allocation. The memory blocks are particularly suited for small-footprint applications and provide sufficient capacity for a range of educational and research scenarios. That being said this separation does result in a larger footprint, primarily due to the extra space required for routing of the separate I/O ports for each bank.

\subsection{Peripheral Enhancements}
The UET-RVMCU includes a comprehensive set of peripherals designed to support typical microcontroller tasks. It provides three general-purpose I/O ports—Ports A, B, and C—with each port featuring 8 pins. These I/O lines can be configured for input or output operation. Additionally, the MCU supports configurable level-sensitive interrupts, enabling rapid response to external events. A GP-Special peripheral module is also integrated to drive 16 LEDs and read from 16 switches, simplifying hands-on experiments and lab exercises.

\subsection{Instruction Set Extensions}
The MCU extends the base RV32IMA instruction set by incorporating the Bitmanip extension, specifically the Zba, Zbb, Zbc, and Zbs subsets. These extensions enhance performance for bit-level operations common in embedded control, signal processing, and data encoding. Although not yet implemented, floating-point instruction support is planned in future versions to accommodate applications requiring higher numerical precision.

\subsection{Debug Support}
To aid in development and system analysis, the UET-RVMCU features built-in debug and trace capabilities. These allow real-time monitoring of internal processor states, support for breakpoints, single-step execution, and program tracing. Such features are invaluable for both educational purposes and professional embedded system debugging.

\begin{figure}[h]
    \centering
    \includegraphics[scale=0.4]{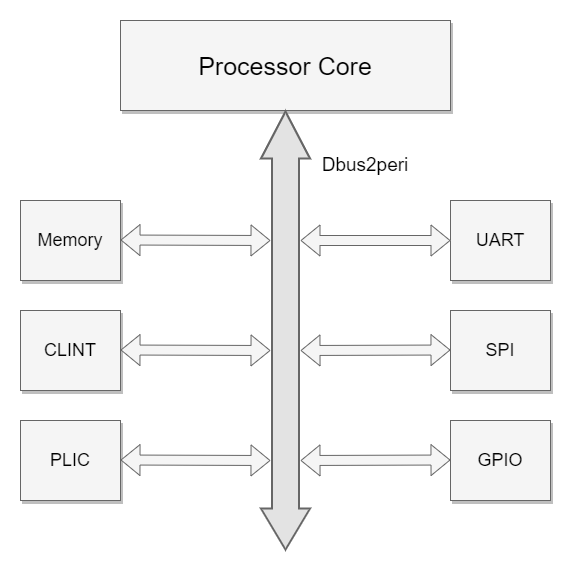}
    \caption{Block diagram of the UET-RV Microcontroller Unit (MCU).}
    \label{fig:uet_rv_mcu_diagram}
\end{figure}

\section{Physical Design Methodology}

\subsection{RTL-to-GDSII Flow}
The physical design of the UET-RVMCU was carried out using the OpenLane toolchain, which provides a complete RTL-to-GDSII flow. The entire flow was based on the open-source Skywater 130nm Process Development Kit (PDK). This methodology allowed us to take the Verilog RTL of our microcontroller design and convert it into a layout-ready GDSII file, suitable for manufacturing.

The flow consists of several key stages:

\subsubsection{Synthesis}
Yosys, an open-source logic synthesis tool, was employed to convert the RTL design into a gate-level netlist. This phase includes optional synthesis exploration to evaluate different area and timing trade-offs. Four default strategies were available to explore the synthesis space. Each strategy was assessed using static timing analysis (STA) to determine optimal results, visualized via an HTML dashboard.

\subsubsection{Floorplanning}
The floorplanning stage used fixed constraints inherited from the Caravel-mini framework. The macro hardening approach guided this stage, allowing us to define precise physical boundaries and pin placements. Custom I/O pin placements were supported, and contextualized I/O placement was used during SoC-level integration to optimize routing.

\subsubsection{Placement}
This phase included both global and detailed placement using tools integrated from the OpenROAD project. The placer positioned standard cells while considering routing congestion and timing requirements. Post-placement optimizations were done using OpenPhySyn, and Clock Tree Synthesis (CTS) was executed to meet timing constraints.

\subsubsection{Routing}
Routing was limited to the Metal-3 layer due to Power Delivery Network (PDN) constraints. The router generated the physical interconnect between cells while ensuring DRC-clean routes. Diode insertion was also done at this stage to prevent antenna effects.

\subsubsection{GDSII Generation}
The final layout was generated using Magic, and GDSII views were produced. KLayout was used to visualize the generated layout. LEF views were also created to support hierarchical design integration.

\subsubsection{Post-routing Evaluation}
Post-routing checks included Design Rule Checking (DRC) and Layout Versus Schematic (LVS) verification using Magic and Netgen. Antenna rule checks were performed via OpenROAD's ARC or Magic. Parasitic extraction using SPEF EXTRACTOR followed, and the extracted data was fed back into STA for accurate timing reports.

\subsection{OpenLANE Infrastructure and Tooling}
OpenLANE is a wrapper framework that integrates tools from OpenROAD, YosysHQ, and Open Circuit Design. The project began in 2019 when most tools were standalone. OpenLANE adopted a more cohesive infrastructure using OpenDB, which allowed a shared physical data model among tools, reducing file I/O overheads and improving iteration efficiency.

Custom tools and utilities built on OpenDB supplemented standard tools to fill in compatibility and methodology gaps. As a result, OpenLANE supported:
\begin{itemize}
\item Hardening designs into soft macros (RTL to GDSII for modules).
\item Integration of macros into complete SoC designs.
\item Support for fully automated and custom interactive flows.
\end{itemize}
The ability to re-harden macros with fixed interfaces made the top-down design methodology viable. Additionally, strict I/O placement specifications allowed for reusable and easily upgradable macros. OpenLANE was instrumental in successfully taping out various RISC-V based SoCs like striVe, showcasing its viability in both academic and industrial contexts.

\subsection{Summary}
The combination of OpenLane, the Skywater 130nm PDK, and open-source tools enabled a robust and fully open-source physical design flow. From RTL synthesis to layout generation, and all the way through verification, this methodology provides a replicable, cost-effective pipeline for academic and pre-production silicon design efforts.

\subsection{Module Integration}
Each SoC component was modularized during floorplanning. The dimensions of each module are listed below:

\begin{table}[h]
\centering
\caption{Module Dimensions}
\begin{tabular}{|l|c|}
\hline
\textbf{Module} & \textbf{Dimensions (\(\mu\)m)} \\
\hline
Core & 700 $\times$ 700 \\
Memory & 1200 $\times$ 1200 \\
DBus & 200 $\times$ 1200 \\
CLINT & 200 $\times$ 100 \\
PLIC & 100 $\times$ 100 \\
GPIO & 200 $\times$ 200 \\
UART & 200 $\times$ 200 \\
SPI & 200 $\times$ 250 \\
\hline
\end{tabular}
\end{table}

\section{Implementation Challenges}

\subsection{Synthesis-Level Constraint with Yosys}
During the physical design implementation using OpenLane, one of the major roadblocks encountered was during the synthesis stage, which internally uses Yosys for hardware description language (HDL) synthesis. Yosys imposes specific syntactic constraints on how reset logic is described. Notably, it does not permit combining the asynchronous reset condition with other signal-based conditions in a single \texttt{if} statement within an \texttt{always} block sensitive to multiple edges.

For instance, take a look at the following:
\begin{figure}[!htb]
\centering
\begin{lstlisting}[style=ieee_verilog, caption={Un-Synthesizable Verilog for an asynchronous reset.}, label={lst:verilog_bad}]
always @(posedge clk or posedge rst) begin
    if (rst || enable) begin
        data_out <= 0;
    end else begin
        data_out <= data_in;
    end
end
\end{lstlisting}
\end{figure}

This will results in the synthesis error:
\begin{figure}[!htb]
\centering
\begin{lstlisting}[style=terminal_output, caption={}, label={lst:terminal_output}]
ERROR: Multiple edge sensitive events found for this signal!
\end{lstlisting}
\end{figure}

This is because Yosys interprets such a condition as a multiple-edge sensitivity problem, which is not supported in its synthesis flow. Even though conceptually the reset and logic signals may seem compatible in a combined condition, Yosys requires that the reset condition be handled independently to avoid ambiguity during synthesis.

The correct approach is to separate the reset logic into its own exclusive \texttt{if} clause, as shown below:
\begin{figure}[h]
\centering
\begin{lstlisting}[style=ieee_verilog, caption={Synthesizable Verilog for an asynchronous reset.}, label={lst:verilog_good}]
// Correct: reset handled in a separate 
// conditional block 
always @(posedge clk or posedge rst) begin
    if (rst) begin
        data_out <= 0;
    end else if (enable) begin
        data_out <= data_in;
    end else begin
        data_out <= data_out;
    end
end
\end{lstlisting}
\end{figure}

This structure ensures that the reset condition is evaluated independently before any other signals, thus complying with Yosys's requirements and avoiding synthesis errors. This debugging insight proved crucial for progressing past synthesis and moving forward to placement and routing.

\subsection{Floorplanning and Congestion During Placement}
Another challenge was encountered in subsequent stages, particularly during floorplanning and standard cell placement. OpenLane's automated flow can sometimes result in uneven cell distribution or congestion hotspots, especially when the design includes modules with irregular aspect ratios or high interconnectivity. In such cases, the global placement engine may struggle to minimize wirelength and timing violations simultaneously.

To address this, manual intervention was required in the form of floorplanning constraint tuning. This involved explicitly defining core utilization, adjusting the die area, and experimenting with different aspect ratios using the \texttt{floorplan.tcl} configuration file. Additionally, pin placements had to be refined using macros such as \texttt{place\_io.tcl} to avoid overlap and improve routability.

These iterative adjustments were necessary to ensure that the downstream stages—global routing, detailed routing, and timing analysis—could proceed without violating design rules or timing constraints. Such manual tuning underscored the importance of understanding both the automated flow and the levers available for optimization in the OpenLane toolchain.

\subsection{Memory Integration Challenges}
Despite the benefits of using separate memory banks for parallelism and modular access, their integration introduced significant complications in our area-constrained design. Each bank had individual read and write ports, which caused severe routing congestion. The numerous input/output pins associated with each memory instance led to overlapping routes and difficulty in achieving Design Rule Check (DRC) clean routing.

The challenge was compounded during global and detailed routing stages, where the tool struggled to find legal paths around the densely connected ports. In future iterations, the use of a shared port memory or serialized access logic may be considered to mitigate this issue.

\subsection{Manual Macro Placement Using Inkscape}
Macro placement also presented difficulties. Initially, we relied entirely on scripted placement provided by OpenLane. However, we discovered that automated placement often led to suboptimal macro distribution and alignment, contributing to routing congestion and poor utilization.

To resolve this, we turned to a manual macro placement strategy using Inkscape, a graphical design tool. We visually designed the macro floorplan layout, ensuring adequate spacing and alignment. Once finalized, we extracted the coordinate information from the Inkscape layout and translated it into the macro placement configuration file required by OpenLane.

This hands-on approach gave us finer control over the floorplan and significantly improved routability.

\subsection{Routing Congestion and Area Utilization}
After macro placements were finalized and the flow was rerun, we encountered significant routing congestion in specific regions. This was verified using the routing heatmap provided by the OpenROAD GUI, which revealed that several areas experienced high congestion while other parts of the die were underutilized.

Interestingly, regions we had intentionally left open for routing were not being effectively used by the tool. This suggested suboptimal routing behavior or a need for better guidance.

To mitigate this, we iteratively updated the macro layout in Inkscape, reallocating space and adjusting positions to balance congestion. After multiple refinements, we observed substantial improvement in route quality and a more even distribution of routing resources across the die.

\section{Results and Discussion}
The final GDSII layout was successfully generated within the 4\,mm\textsuperscript{2} area constraint. The design’s modular nature allows for easy integration of additional peripherals, supporting future scalability.

This project highlights the experience of using a fully open-source EDA toolchain, making it ideal for education and prototyping. The micro-controller is simple yet functional, offering value for student learning and basic embedded applications.

However, the design currently lacks integrated SRAM and does not yet include a floating-point unit or debug support, which are planned for future improvements.

\begin{figure}[h!]
    \centering
    \includegraphics[width=0.45\textwidth]{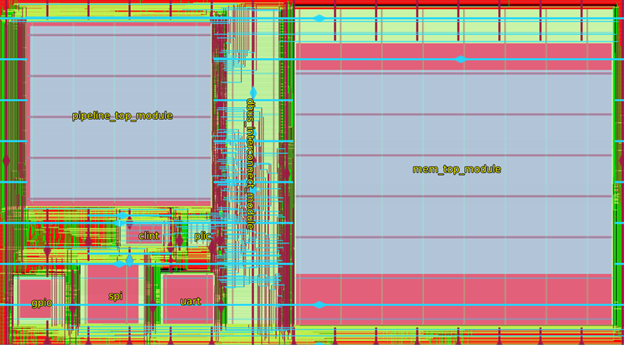} 
    \caption{Final GDSII layout of the SoC}
    \label{fig:gdsii}
\end{figure}

\section{Conclusion and Future Work}
The UET-RVMCU demonstrates a successful open-source implementation of a lightweight RISC-V microcontroller derived from an application-class processor. Its simplified architecture supports compact embedded designs. Future work includes adding floating-point hardware support, enabling full debug and trace capabilities, and evaluating silicon performance post-fabrication through shuttle runs. There is also the prospect of adding a vector coprocessor to enable low-power, edge AI applications.

\end{document}